\documentclass[journal]{IEEEtran}
\usepackage{amsmath}
\usepackage{amsthm}
\usepackage{xfrac}
\usepackage{amssymb}
\usepackage{mathabx}
\usepackage{psfrag}
\usepackage{graphicx}
\usepackage{epstopdf}
\usepackage{todonotes}
\usepackage{cite}
\usepackage{algorithmic}
\usepackage{algorithm}
\usepackage{svg}
\usepackage{mathtools}
\usepackage{hyphenat}
\usepackage{booktabs}
\usepackage{accents}
\usepackage{bbm}
\usepackage{balance}
\usepackage{verbatim}
\usepackage{subcaption}
\newlength{\dhatheight}

\usepackage{float}
\usepackage{multirow}

\usepackage{xparse,mathtools}

\graphicspath{{"figures/"}}
\DeclareGraphicsExtensions{.pdf,.jpg,.png}

\begin{document}

\title{Mobile Distributed MIMO (MD-MIMO) for NextG: Mobility Meets Cooperation in Distributed Arrays}
\author{Karim A. Said, Yibin Liang, Usama Saeed, Ramin Safavinejad, Kumar Sai Bondada, Evan Allen, Nima Mohammadi, Benjamin Pimentel, Brian Kelley, Jeff Reed, Nishith Tripathi, Michael Buehrer, Yang (Cindy) Yi, Daniel Jakubisin, and Lingjia Liu.
\thanks{K. Said, Y. Liang, U. Saeed, R. Safavinejad, K. S. Bondada, E. Allen, N. Mohammadi, J. Reed, N. Tripathi, M. Buehrer, Y. Yi, D. Jakubisin, and L. Liu are with Wireless@Virginia Tech, the Bradley Department of ECE at Virginia Tech, Blacksburg, VA, USA. B. Pimentel is with Naval Postgraduate School, USA. B. Kelley is with the Electrical and Computer Engineering Department, University of Texas at San Antonio, San Antonio, TX, USA. The corresponding author is L. Liu (ljliu@ieee.org).}
\thanks{Efforts sponsored by the U.S. Government under the Training and Readiness Accelerator II (TReX II), OTA. The U.S. Government is authorized to reproduce and distribute reprints for Governmental purposes notwithstanding any copyright notation thereon.

The views and conclusions contained herein are those of the authors and should not be interpreted as necessarily representing the official policies or endorsements, either expressed or implied, of the U.S. Government.}
}

\maketitle
\begin{abstract}
Distributed multiple-input multiple-output (D\mbox{-}MIMO) is a promising technology to realize the promise of massive MIMO gains by fiber-connecting the distributed antenna arrays, thereby overcoming the form factor limitations of co-located MIMO. 
In this paper, we introduce the concept of mobile D-MIMO (MD-MIMO) network, a further extension of the D-MIMO technology where distributed antenna arrays are connected to the base station with a wireless link  allowing all radio network nodes to be mobile.
This approach significantly improves deployment flexibility and reduces operating costs, enabling the network to adapt to the highly dynamic nature of next-generation (NextG) networks.
We discuss use cases, system design, network architecture, and the key enabling technologies for MD-MIMO. Furthermore, we investigate a case study of MD-MIMO for vehicular networks, presenting detailed performance evaluations for both downlink and uplink. The results show that an MD-MIMO network can provide substantial improvements in network throughput and reliability.
\end{abstract}



\section{Introduction}
\looseness=-1
Multiple-input multiple-output (MIMO) is an enabling technology for 4G networks that significantly increases network capacity along with improvement in reliability~\cite{liu2012downlink}.
As an extension, massive MIMO has been introduced in 5G networks to further enhance various network performance metrics.
However, the form factor limitation at the base station typically restricts the number of antenna elements, which in turn limits the achievable network performance.  
Accordingly, distributed MIMO (D-MIMO) operation where multiple radio units (RUs) cooperate to assist a given node in either the transmission of information to destination node(s), or the reception of information from source node(s), has been introduced.
This concept, which originated from coordinated/cooperative multi-point transmission (CoMP) in the 4G era, has been refined through techniques such as CoMP joint transmission using channel information feedback and higher rank dedicated beam‐forming~\cite{liu2013coordinated} for cooperative communications, and has been adopted into the 5G NR standards in the form of multiple Transmission and Reception Point (mTRP)~\cite{chen20235g}. 


The key advantage of distributed MIMO (D-MIMO) over co-located MIMO is its ability to provide flexible MIMO operation in terms of
joint transmission~\cite{liu2012system} and interference coordination~\cite{liu2012inter} while overcoming the form factor limitation at the base station.
In the context of 5G networks, RUs capable of coherent joint transmission can operate as a virtual antenna array with increased spatial multiplexing and higher antenna diversity.
Less sophisticated D-MIMO strategies such as Distributed Antenna Systems (DAS) and Dynamic Point Selection (DPS) can rely on macro diversity to achieve coverage gains due to the geographical distribution of the RUs. 

Most existing D-MIMO systems are based on fixed infrastructure networks, such as 5G/6G terrestrial networks. Given the highly dynamic nature of modern networks in terms of demand and coverage, modifying the fixed infrastructure to adapt to such changes is both expensive and time-consuming. Although non-terrestrial network (NTN) can cover large areas, they fail to provide the same bandwidth and coverage density as terrestrial systems. To extend the benefits of D-MIMO into more dynamic environments, node mobility becomes essential, enabling improved network adaptability, resource utilization, and service continuity.

In support of 6G’s vision for ubiquitous and intelligent connectivity, flexible and intelligent distributed architectures are becoming increasingly vital. Recent work on Distributed Intelligent Sensing and Communications (DISAC) and Extremely Large-Scale MIMO (XL-MIMO) highlights the role of distributed MIMO systems in enabling real-time sensing, semantic data processing, and adaptive coverage, while also addressing challenges in distributed signal processing, synchronization, and energy efficiency, all of which are critical enablers for next-generation wireless networks  \cite{DISCAC_wirelesscomms},  \cite{XLMIMO_wirelesscomms}.

\begin{figure}[ht]
  \centering
  \begin{subfigure}{\columnwidth}
  \centering
    \includegraphics[width=1\linewidth]{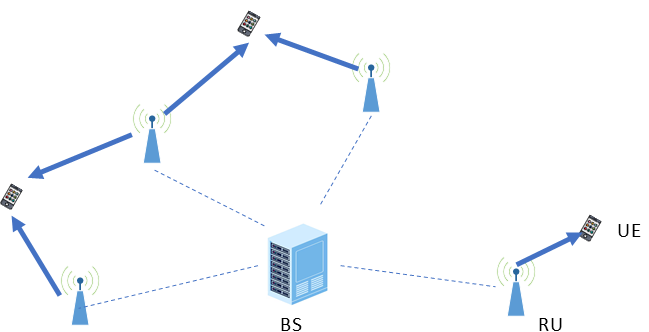}
    \caption{Traditional D-MIMO Architecture}
    \label{DMIMO_DL}
  \end{subfigure}
  \begin{subfigure}{\columnwidth}
  \centering
    \includegraphics[width=1\linewidth]{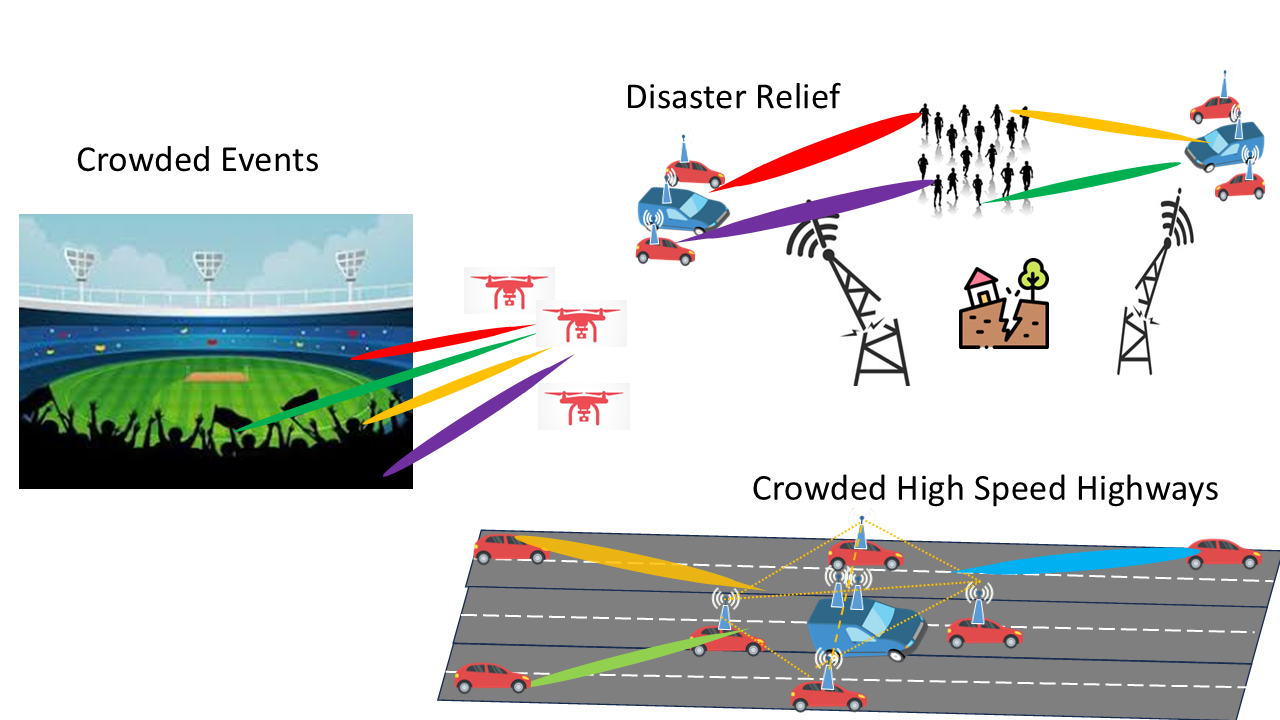}
    \caption{MD-MIMO: Node mobility as a key enabler for diverse application scenarios.}
  \label{app_sc}
  \end{subfigure}

    \caption{D-MIMO versus MD-MIMO}

  \label{MD_MIMO_arch}
  
\end{figure}

We refer to mobile distributed MIMO (MD-MIMO) as an evolution of D-MIMO that allows all nodes, including the gNB, to be mobile or portable, thereby enabling full control over both micro- and macro-diversity. As illustrated in Figure \ref{app_sc}, MD-MIMO can address critical scenarios such as providing responsive connectivity during disaster relief, enhancing capacity at crowded events, and extending coverage for vehicular or UAV communications.

We identify the main differences between conventional D-MIMO and MD-MIMO in Section \ref{sec:dmimo2mdmimo}. In Section \ref{sec:dmimo-arch}, we explore various architectures for an MD-MIMO system, and Section \ref{sec:enabling-tech} addresses the challenges and technologies that make such systems viable. Section \ref{sec:simulation} provides some preliminary results for vehicular MD-MIMO uplink and downlink use cases. Finally, Section \ref{sec:conclusions} summarizes current insights and outlines future research direction.

\section{From D-MIMO to MD-MIMO} \label{sec:dmimo2mdmimo}

\subsection{Wireless Front-haul}

Figure \ref{DMIMO_DL} illustrates a typical downlink D-MIMO layout. A base station (gNB) is linked to multiple nodes (RUs) which jointly act as a transmit array to serve several users. The gNB-to-RU link is a high-speed wired connection, known as the front-haul (FH), typically implemented with optical fibers. The FH link carries high-speed data streams containing physical-layer IQ samples between the gNB and the RUs, enabling the offloading of baseband processing to the gNB.  

Wired FH links tether each RU to existing infrastructure, rendering the network quasi-static. 
Although microwave links offer wireless connectivity, the high directionality inherent in microwave communications makes them unsuitable for mobile RUs. This inflexibility of FH link limits network optimization through RU relocation or the addition of new RUs.
Freed from the limitations of rigid FH links, networks can adapt to dynamic load conditions with unprecedented flexibility. 
Leveraging this advantage, dynamic networks maintain service and coverage in fundamentally different ways compared to static networks. For example, a UAV-based RU could adjust its altitude and position over a crowded public space to establish an LOS link with a greater number of users, while a vehicular RU might create a mobile bubble of connectivity to nearby vehicles. 

Advancements in the computational capabilities of modern UEs enable them to take on an altruistic role, acting as RUs to facilitate transmission and reception between the gNB and other UEs. The available UEs can be opportunistically utilized to supplement traditional network resources in high-demand circumstances. 
By efficiently selecting UEs in favorable conditions, the network enhances performance during peak times while avoiding the periods of inactivity typically associated with dedicated wired RUs during unfavorable channel conditions--thereby improving overall network efficiency.


\subsection{Implications of Mobility}

A network which utilizes a distributed antenna system composed of a portable gNB and portable RUs can address a variety of use cases, thereby enhancing network performance, coverage, and resource utilization. As shown in Fig. \ref{app_sc}, the independence of an MD-MIMO system enables rapid deployment at disaster relief sites.
The system supports scenarios with periodic high demand, such as sports events or concerts, without requiring permanent infrastructure deployment.
Furthermore, in high-mobility use cases, such as high-speed traffic, MD-MIMO can prevent frequent handovers. However, when both the gNB and the RUs are mobile, the wireless FH link channel becomes more dynamic and challenging. The non-uniformity and constant variability of the virtual array pattern will have significant implications on the channel state information (CSI) feedback and precoding. Furthermore, high mobility speeds significantly increase the complexity and frequency of RU calibration, as the required calibration rate scales with the number of nodes (gNBs and RUs) and their velocities.  Movement directly impacts propagation delay calibration and the accuracy of channel estimates, while distinct mobility-induced Doppler shifts at each antenna further complicates the synchronization and calibration process.




\subsection{Coherence}

D-MIMO cooperation strategies can be categorized according to the transmission techniques employed. Coherent Joint Transmission (CJT) enables nodes within a D-MIMO network to synchronize to a common phase and timing reference. This synchronization allows a group of nodes to create virtual arrays for transmission or reception. Such configurations can emulate the performance of much larger co-located arrays, offering significant enhancements in signal diversity and cost efficiency. For coherent transmission, timing, frequency, and phase synchronization are required across all nodes. Achieving this level of synchronization incurs additional overhead signaling across the wireless FH link.  

Coherent transmission depends on the synchronization of the transmitted waveform from each RU; these waveforms must constructively interfere at the receiver. To achieve this, each RU must maintain a synchronized frequency reference signal, ensuring that the frequencies and phase alignments assumed by the spatial precoder are accurate. Once established, synchronization must be maintained and periodically adjusted throughout the transmission, with the update frequency determined by hardware capabilities and channel conditions.

Conversely, Non-Coherent Joint Transmission (NCJT) can alleviate the intensive synchronization overhead  characteristic of CJT schemes. 
However, this reduction in overhead complexity compromises the nodes' ability to form virtual arrays that approximate the performance of their co-located counterparts. Despite this limitation, nodes can still employ cooperative strategies to enhance performance. Depending on available resources and specific demands, MD-MIMO provides a range of adaptable procedures tailored to best suit each scenario. 

\subsection{Distributed Processing}
\looseness=-1
Unlike the wired FH links, wireless FH link cannot be treated as an error-free medium and must be used efficiently given the capacity limitations and inherent instability of a wireless channel. In a wired D-MIMO system, typical RUs exchange sample-level information with the gNB--where all processing functions reside--and are limited to basic functionality such as performing frequency up- and down-conversion. In contrast, for a wireless FH, it is more practical to exchange only symbol-level information. Consequently, signal processing must be distributed among the RUs so that only essential information is exchanged. This approach requires each RU to have full transceiver functionality, including channel estimation and MIMO precoding. Moreover, the distance between the RUs and the source gNB must be carefully taken into account to ensure sufficient FH link capacity. This trade-off means that while RUs located farther from the source can enhance macro-diversity, they may suffer from reduced throughput due to lowered FH capacity.





\section{MD-MIMO Architecture} \label{sec:dmimo-arch}


In an MD-MIMO system similar to that shown in Fig. \ref{DMIMO_DL}--but with all nodes mobile and connected wirelessly--a complete transmission from source to destination occurs in at least two phases in time. The first phase (or ``hop") is from the gNB to the RUs, and the second is from the RUs to the destination UE. Within each  hop, multiple wireless links must be managed to minimize interference. Factors affecting interference include the number of destination nodes, the placement and distribution of nodes with wrieless FH links, and the number of spatial streams. We consider two cooperation architectures for the two corresponding communication directions--a transmit virtual array for downlink and a receive virtual array for uplink--with the design implications of the uplink-downlink asymmetry explored in each case.

\begin{figure}[ht]
  \centering
  \begin{subfigure}{\columnwidth}
  \centering
    \includegraphics[width=0.85\linewidth]{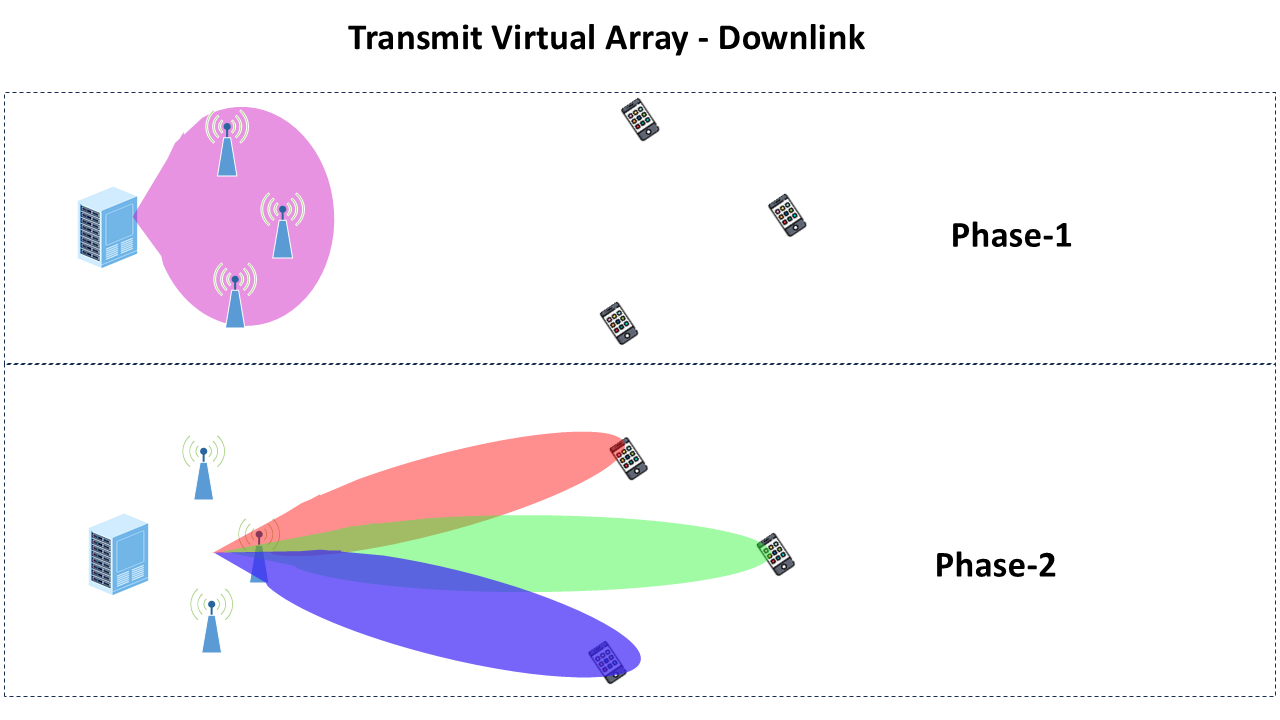}
   \caption{Transmit Virtual Array - Downlink} \label{tx_arr}
  \end{subfigure}
  \begin{subfigure}{\columnwidth}
  \centering
    \includegraphics[width=0.85\linewidth]{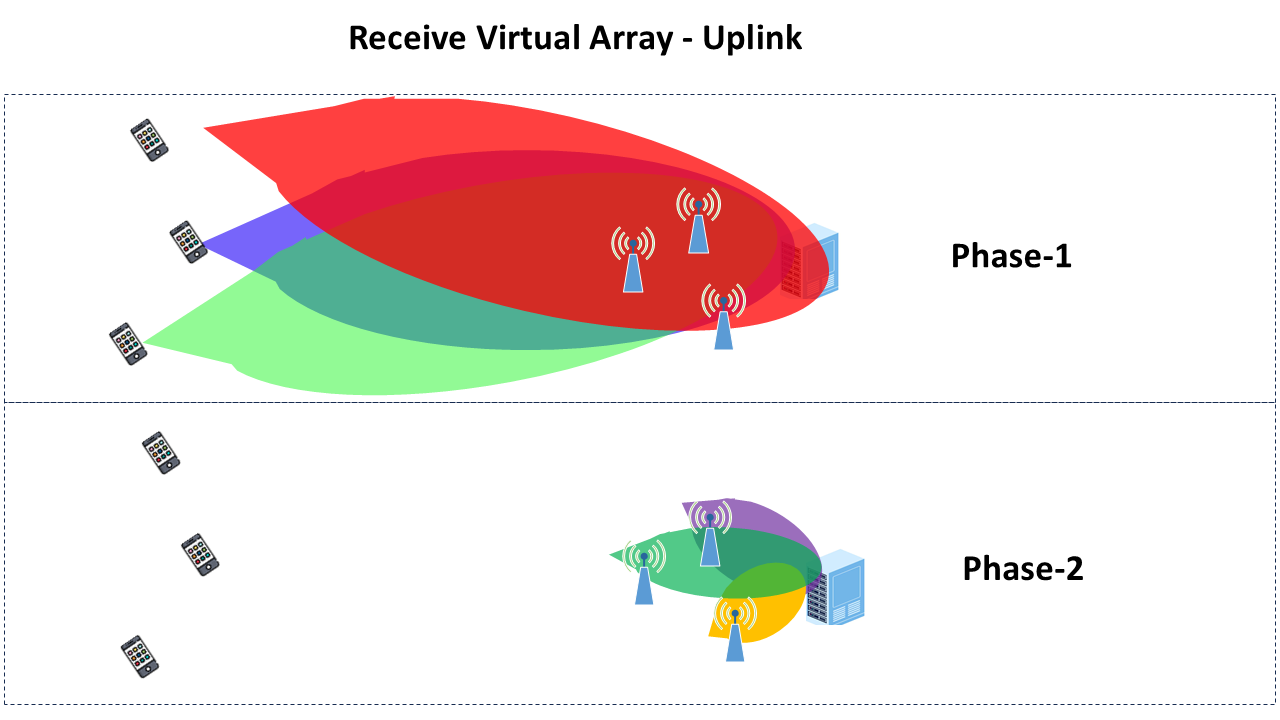}
\caption{Receive Virtual Array - Uplink} \label{rx_arr}
  \end{subfigure}
    \caption{MD-MIMO Topologies}
  \label{MD_MIMO_topology}
\end{figure}

\subsection{Transmit Virtual Array}
In the downlink architecture, the gNB together with a cluster of nearby RUs forms a virtual array that transmits signals to destination UEs. This virtual array enhances reliability by leveraging both the aggregate power of multiple distributed transmitters and transmit diversity. When channel state information is available and phase coherence is maintained among the virtual array nodes, spatial multiplexing gains can be realized. 
\subsubsection{Single UE destination node}
In the case of a receiving UE experiencing poor direct link conditions, such as edge users, the virtual array can offer significant reliability gains and extended range. However, throughput is ultimately limited by the number of antenna elements at the destination UE.

Transmission occurs in two phases. 
In the first phase, the source gNB broadcasts a common data set to all RU nodes (as illustrated by the purple beam in Fig. \ref{tx_arr}. For a typical gNB
with multiple antennas, a precoder is used to send multiple streams of information that will all arrive at all RUs. It is ideal that all nodes forming the virtual array have
identical copies of the total information to be transmitted.
In the second phase, the channel corresponding to the link between the virtual array and the destination UE is a MIMO channel which can support 2 streams at most when the destination UE has two antennas. A conventional  MIMO precoder may then be employed.  



\begin{figure}[ht]
  \centering
  \begin{subfigure}{\columnwidth}
  \centering
    \includegraphics[width=0.7\linewidth]{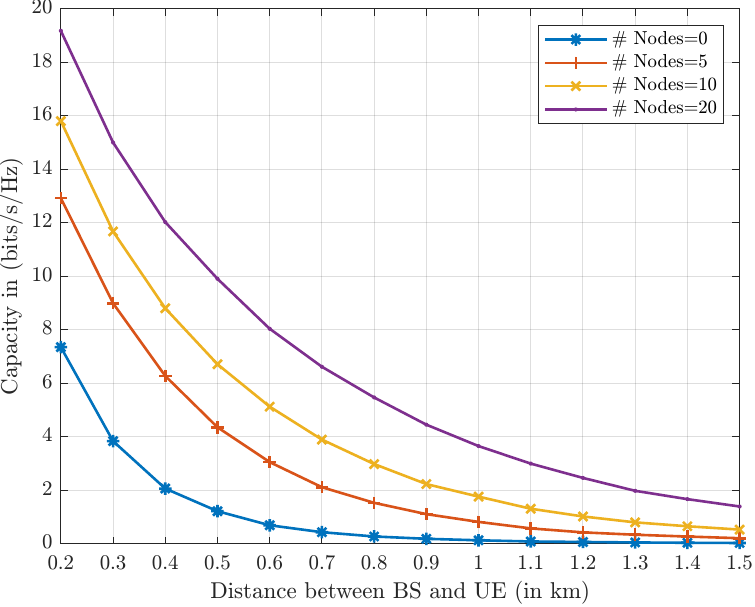}
    \caption{Phase-2 Capacity plots}
    \label{fig:inter_squad_dist}
  \end{subfigure}
  \begin{subfigure}{\columnwidth}
  \centering
    \includegraphics[width=0.7\linewidth]{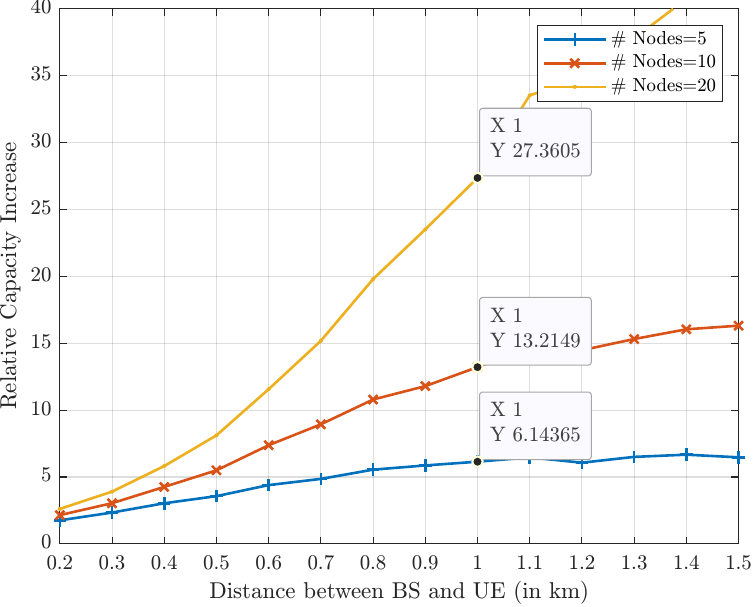}
    \caption{Phase-2 Capacity Relative increase over the baseline}
    \label{fig:range_improve}
  \end{subfigure}
    \caption{Average capacity and relative capacity (multiplicative) increase over the baseline using ZF precoding under UMi settings.}
  \label{arch1_results}
  
\end{figure}

Fig. \ref{arch1_results} presents performance plots of the spectral efficiency achieved versus the distance between the source and destination node in urban micro settings (see \cite{10757642} for detailed analysis). In this setup, the source node is a base station (BS) with four transmit antennas, and the receiver is a UE with two receive antennas. The RUs in this configuration, representing other users in the network, coherently transmit jointly with the BS using a zero-forcing (ZF) precoding strategy. The transmission powers for the BS and RUs are set to $33$ dBm and $26$ dBm, respectively. The results demonstrate that the capacity is significantly improved, with higher relative improvements observed at greater distances. The relative increase in capacity is the capacity improvement achieved over the baseline, where the BS communicates directly with the destination UE, and the plots show relative (multiplicative) increments of $27.36$, $13.21$, and $6.14$ times improvement at $1$ km distance.

\subsubsection{Multiple UE destination nodes}
When the gNB sends information to serve multiple destination UEs, the spatial dimensionality of the distributed array is exploited to multiplex several streams of information to different UEs. In the first phase, the total information set intended for all receiving UEs is broadcast to every node participating in the virtual array. In the second phase, spatial multiplexing is achieved using multi-user MIMO (MU-MIMO) precoding techniques to guide spatial streams to their corresponding receiver UEs. The adoption of standardized codebooks for precoding can improve the interference avoidance capability further by enabling the dynamic selection of transmission directions that minimize interference~\cite{liu2012inter}. Alternatively, a combination of spatial multiplexing and successive interference cancellation (SIC) techniques, as in rate splitting multiple access (RSMA), can be used to minimize inter-stream interference. 

In contrast to D-MIMO with wired FH, having a cascade of two wireless links (gNB-to-RU, RU-to-UE) suggests that the optimal transmission strategy might require joint optimization over the two links. For example, from the point of view of the max-min precoder strategy for the phase-1 link, omitting certain RUs may improve the achievable rate; conversely, in the phase-2 link, incorporating additional RUs increases the spatial dimensionality of the virtual array, thereby enhancing the rate. Thus, the achievable rate exhibits a convex dependence on the number of active RUs in the virtual array.

\subsection{Receive Virtual Array}
For uplink communication, Fig. \ref{rx_arr} illustrates an MD-MIMO architecture in which the gNB, together with multiple RU nodes in close proximity, form a virtual array capable of MU-MIMO reception from several independent source UEs.
Unlike co-located arrays, MIMO combining at the gNB can only be performed if it receives sample-level information from all RU nodes.
However, forwarding sample information from RU nodes to the gNB over the wireless front-haul reduces the effective dimensionality of the virtual array (total number of antennas in the virtual array) down to the number of antennas of a single node, which is the gNB node. 

To circumvent this loss in dimensionality if sample forwarding is used, a \textit{non-coherent} forwarding approach is deemed more efficient. In this strategy, each RU independently decodes information and forwards the decoded information to the gNB without employing precoding at the transmit UEs. 
While this approach eliminates the need for phase-level coherence among the virtual receive array nodes, it still demands symbol-level synchronization. Consequently, each RU must independently separate and decode the multiple streams it receives.

In the first phase of uplink transmission, each source UE sends a data stream that is received by all RU nodes and the gNB. SIC is then applied at each node to decode the incoming streams. In the second phase, the RUs forward the decoded data to the gNB, which uses SIC to further decode the transmissions from the different RUs before performing post-detection fusion.


\section{Enabling Technologies}\label{sec:enabling-tech}

Realizing the potential of MD-MIMO's demands advanced technological solutions to key challenges.
These technological advances are critical for supporting the evolution of wireless networks as envisioned in 5G-Advanced and 6G~\cite{chen20235g}.
In the following subsections, we examine some key challenges and the corresponding enabling technologies. 

\subsection{Distributed Synchronization Algorithms}
One of the key challenges facing MD-MIMO is the synchronization. For coherent MIMO transmission, each RF chain, per antenna, must maintain the same carrier frequency, sampling clock frequency, and transmission timing. Simulation results indicate that achievable MIMO capacity gains are highly sensitive to carrier frequency and clock synchronization offsets. Moreover, beamforming requires additional phase alignment across the antenna array~\cite{mudumbai2009}. 
This uniformity ensures  that the frequency and phase assumptions used by the spatial precoder are accurate across nodes, thereby guaranteeing correct phase alignment at the receiver.

For non-coherent transmission, the strict synchronization required to maintain sample-level coherence among distributed nodes is relaxed. 
However, symbol level synchronization is still needed to enable cooperation among distributed antenna arrays, besides accurate carrier frequency synchronization among simultaneously transmitting nodes. 


In MD-MIMO, each RU must adopt a coordinated frequency reference signal. Adjustment of the frequency synchronization is necessary whenever the center frequency changes. Additionally, as the local oscillators experience drift over time, the compensation factors that were initially established among the RUs will also shift. These factors must be adjusted accordingly to maintain accurate frequency synchronization. 

Once frequency synchronization is achieved, it is crucial that each RU initiates transmission simultaneously. Inaccurate timing can lead to unique perceived phase rotations among transmit-receive antenna pairs due to asynchronous transmissions. 
Therefore, each RU must maintain not only a stable clock but also a common reference to determine the transmission start and end times. 
Distributed timing synchronization may also require propagation delay calibration; by accounting for the transmission delays, the resulting phase rotation can be compensated. 
In scenarios lacking such calibration, rapid channel estimation techniques can be employed to compute an effective channel that incorporates these delays.

Although the effects of relative clock skew generally occurs at a slower rate than that of frequency synchronization, mobility significantly impacts stability in timing. Movement by either RUs or UEs alters propagation delays, rendering previous calibrations inaccurate. Consequently, timing synchronization must be continuously adjusted to maintain proper alignment.

Standard procedures for distributed synchronization typically involve: 
1) achieving local oscillator (LO) clock synchronization among nodes with respect to a reference source;
2) performing iterative timing and phase synchronization among transmitting nodes for coherent transmission; and
3) implementing open-loop or closed-loop synchronization for virtual phased array beamforming. 
Distributed synchronization algorithms are proposed in \cite{nanzer2021} for distributed phased arrays. 

Furthermore, mobility of the RUs adds additional complexity to synchronization requirements. When the gNB and the UEs are moving relative to each other, extra measures must be taken. For instance, Doppler shifts must be compensated for in carrier frequency-related operations, and phase shifts must account for the relative distance and speed of all nodes.

\subsection{ML-based Channel Estimation and Prediction}
\looseness=-1
MD-MIMO systems--with their large number of total antenna elements--pose significant challenges for channel estimation and feedback. 
Although 3GPP standards prescribe procedures for downlink channel estimation and codebook-based CSI feedback to the gNB~\cite{3gpp_38.331} to reduce overhead, the necessary codebook size grows exponentially with the number of transmit antennas
, making frequent CSI feedback impractical. 

Various machine learning-based techniques have been proposed to address these challenges, such as CSI compression, 
optimized codebook design,
 and channel prediction \cite{channel_prediction_TCOM}.
Although these methods may prove effective, they rely on extensive offline training, i.e. machine learning models are first trained on large datasets and then deployed with fixed parameters during (online) system operation.
Such an approach presumes that the offline data accurately mirrors real-time channel conditions; any statistical discrepancy can significantly impair performance \cite{uncertainty_in_generalization}.

To address these challenges, there is a growing need for machine learning techniques that can effectively reduce CSI feedback overhead in an online, adaptive manner.
A particular type of recurrent neural networks (RNNs), namely reservoir computing (RC), has recently emerged as an effective way to tackle problems in wireless communications in an online and real-time manner. RC based methods have been introduced for wireless communication problems such as symbol detection~\cite{RC_struct}, where the models have been shown to be effective at learning spatio-temporal correlations. The RC network consists of an RNN-based reservoir with fixed weights and a trainable output layer.
The light-weight training makes RC highly adaptable to dynamic environments, making it a strong candidate for online channel prediction within MD-MIMO systems. 


\section{MD-MIMO Case Studies}\label{sec:simulation}
\subsection{Coherent MD-MIMO System: Transmit Virtual Array for Downlink}

\begin{figure}[ht]
  \centering
  \begin{subfigure}{\columnwidth}
  \centering
    \includegraphics[width=1\linewidth]{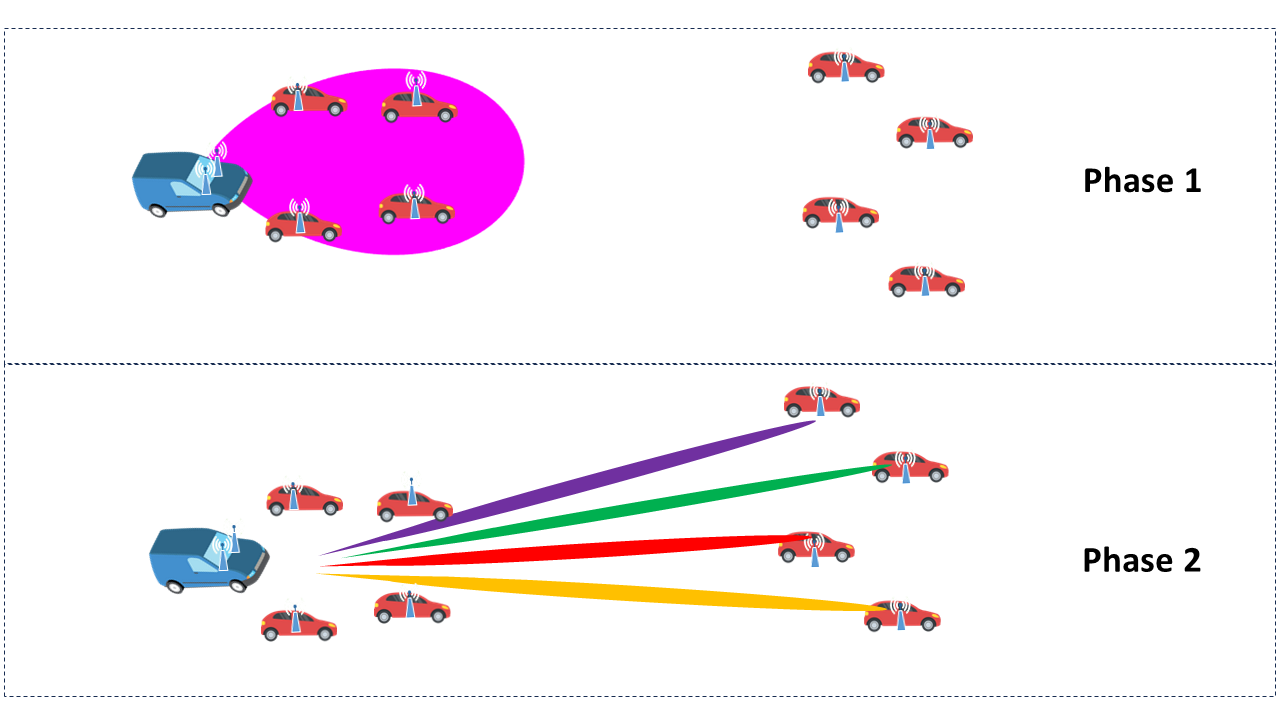}
\caption{Coherent MD-MIMO: Transmit virtual array sending to multiple receivers} \label{single_arry_mult_rx}
  \end{subfigure}
  \begin{subfigure}{\columnwidth}
  \centering
 \includegraphics[width=\linewidth]{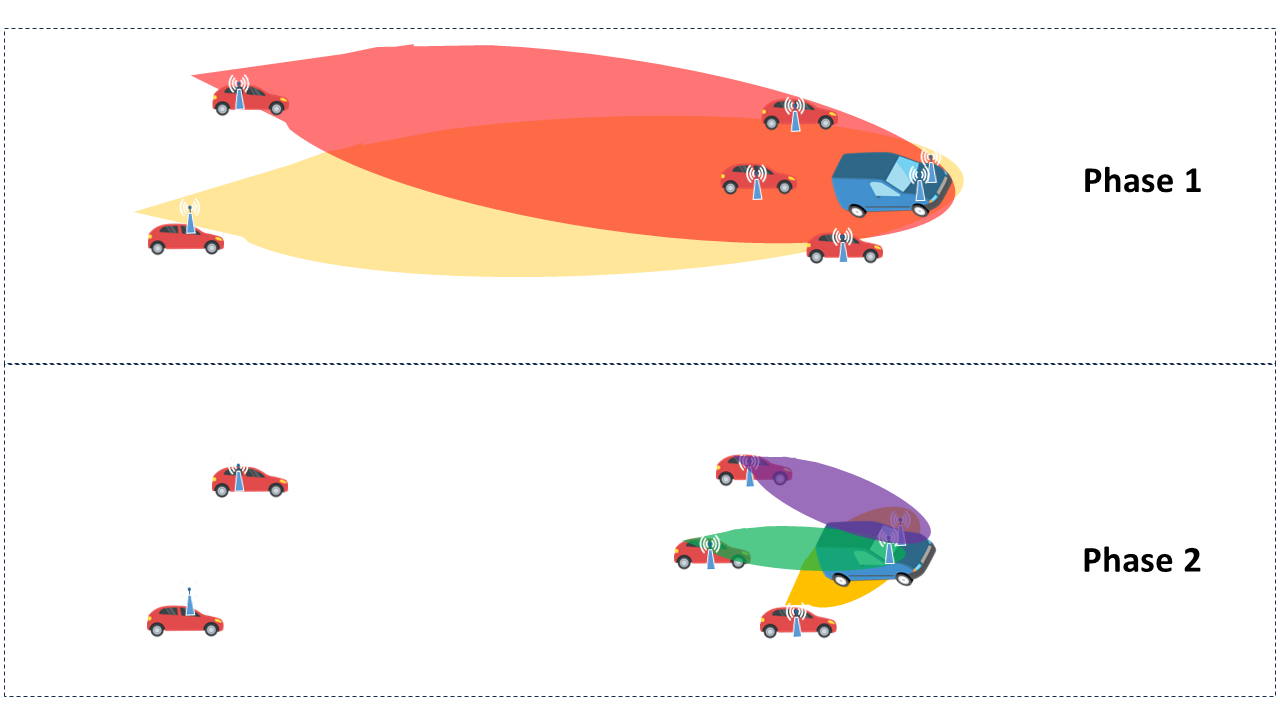}
\caption{Non-coherent MD-MIMO: Multiple independent transmistters sending to receive virtual array} \label{mult_tx_arry_angl_rx_arry}
  \end{subfigure}

    \caption{MD-MIMO Case Studies}

  \label{MD_MIMO_case_study}
  
\end{figure}

We consider a scenario where a vehicle-mounted gNB communicates with multiple mobile UEs. The gNB coordinates with nearby RUs to form a virtual array that enables joint transmission to the destination UEs. Together, the source gNB and neighboring RUs transmit unique data streams to multiple independent receivers as a unified virtual array.

To accurately model mobile distributed channels, we adopt the ns-3 network simulator, leveraging stochastic geometry-based techniques to simulate scenarios in accordance with specifications in 3GPP TR 37.885, which defines the 3GPP vehicle-to-everything (V2X) channel model and addresses the dynamic and high-mobility scenarios of vehicular communication, capturing time-varying Doppler effects, interference, and vehicular-specific LOS/NLOS transitions \cite{3gpp_37.885}. By leveraging stochastic geometry, ns-3 enables realistic simulations of node placement, interference patterns, and spatial consistency, capturing both temporal and spatial correlations.

We use the following parameters to simulate the given scenario: carrier frequency $f_c = 3.5$ GHz, subcarrier spacing $\Delta f = 15$ kHz, total number of subcarriers $N_c = 512$, the gNB's transmision power limit is $35$ dBm, and each RU's transmission power limit is $26$ dBm.  Each gNB is assumed to have 4 antennas and each RU has 2 antennas. We use the 5G NR V2V channel model for the links between all TX-RX antenna pairs. Three different mobility scenarios are considered for evaluating the throughput and spectral efficiency improvement. Table \ref{tbl:mobility-scenario} lists the ground speed of the gNB and the UE speed relative to the gNB for three mobility scenarios. The RC-based channel prediction algorithm is used to improve the precoding performance for the distributed MIMO channels. 

\begin{table}[ht]
    \centering
    \begin{tabular}{c|c|c}
         \hline
         Mobility & gNB ground speed  & UE relative speed \\
         \hline
         Low &  0.1 km/h  & 0.01 km/h \\
         Medium & 3 km/h & 0.3 km/h \\
         High & 10 km/h & 1 km/h \\
         \hline
    \end{tabular}
    \caption{Mobility scenarios for performance evaluation}
    \label{tbl:mobility-scenario}
\end{table}

\begin{figure}[ht]
  \centering
  \begin{subfigure}{\columnwidth}
  \centering
    \includegraphics[width=0.9\linewidth]{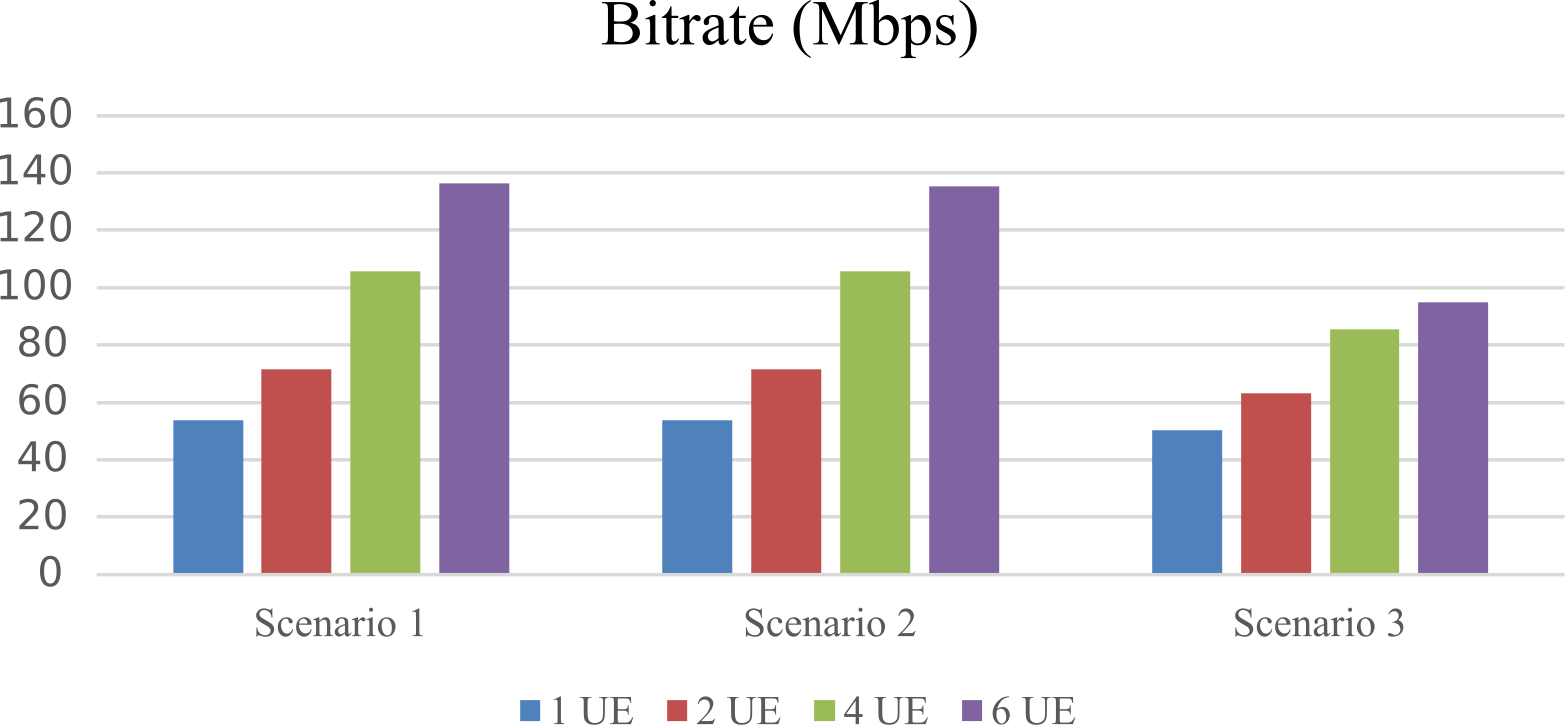}
\caption{Bit Rate in Mbps} \label{mult_tx_mult_rx_tput_results}
  \end{subfigure}
    
    \vspace{20pt}
    
  \begin{subfigure}{\columnwidth}
  \centering
 \includegraphics[width=0.9\linewidth]{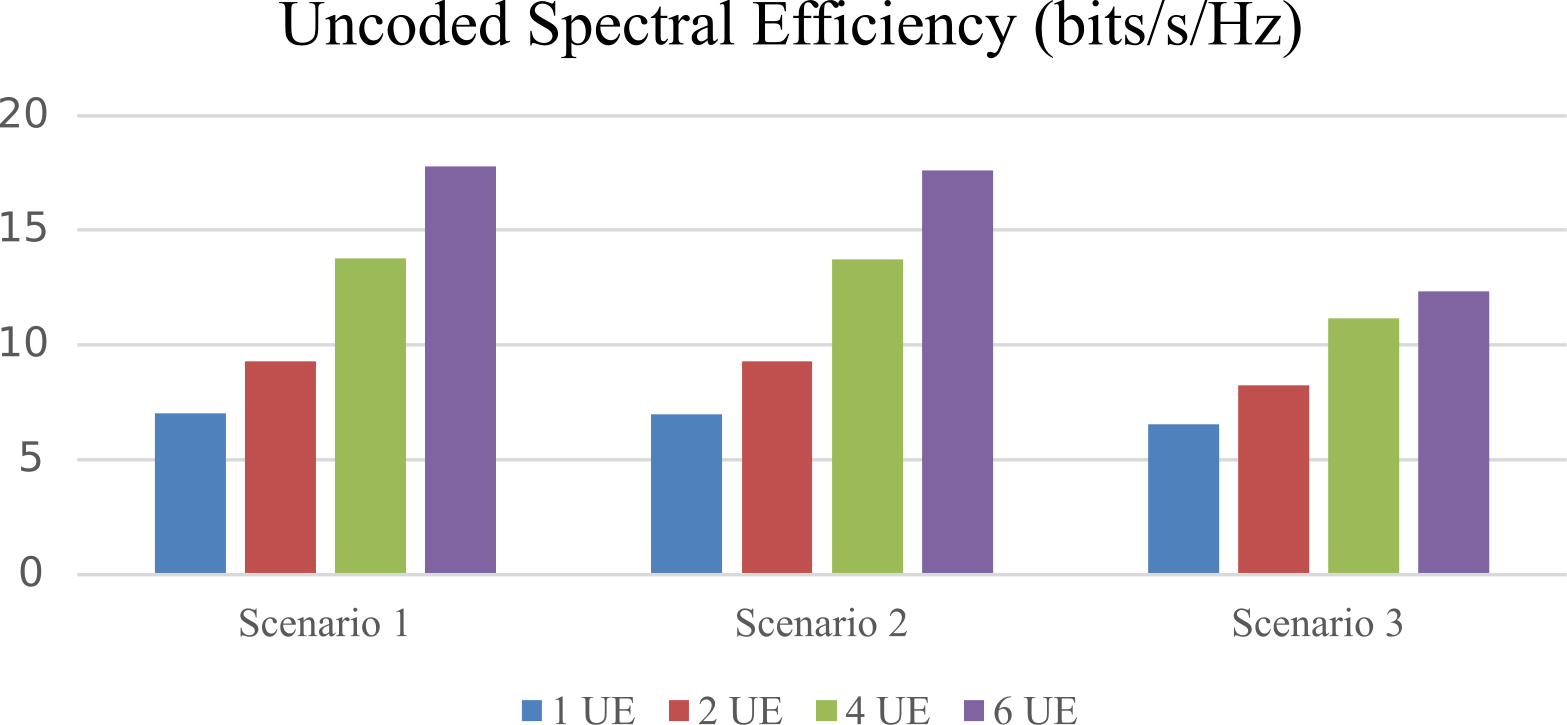}

\caption{Uncoded Spectral Efficiency} \label{mult_tx_mult_rx_se_result}
  \end{subfigure}

    \caption{Bitrate and Uncoded Spectral Efficiency}

  \label{MD_MIMO_results}
  
\end{figure}

Figures \ref{mult_tx_mult_rx_tput_results} and \ref{mult_tx_mult_rx_se_result} show the system throughput and spectral efficiency
results for three scenarios each with a different number of UEs. As the number of UE increases, the overall throughput and
spectral efficiency improve continuously. Comparing high-mobility to low-mobility scenarios, the decrease in throughput is not significant, meaning the distributed RUs can provide performance gains as a virtual array.

\subsection{Non-coherent MD-MIMO Receive Virtual Array for Uplink}
We consider an uplink scenario where two UEs, each equipped with two antennas, transmit their data to the receive virtual array.
In this scenario, each receive node encompasses two antennas and is therefore capably of separating up to two streams.
Given that the UEs collectively possess more than two antennas, they have the option to employ Space Time Block Coding (STBC) to enhance the diversity gains. 
In practical terms, each node within the virtual array receives the signal as each UE encodes its data with STBC before transmitting over the air.
Consequently, the signals from the UEs will interfere with one another, and each receiving node must apply multi-stream STBC decoding mechanisms to distinguish and recover the two independent streams.

\begin{figure}[th]
\centering 
\includegraphics[width=0.9\linewidth]{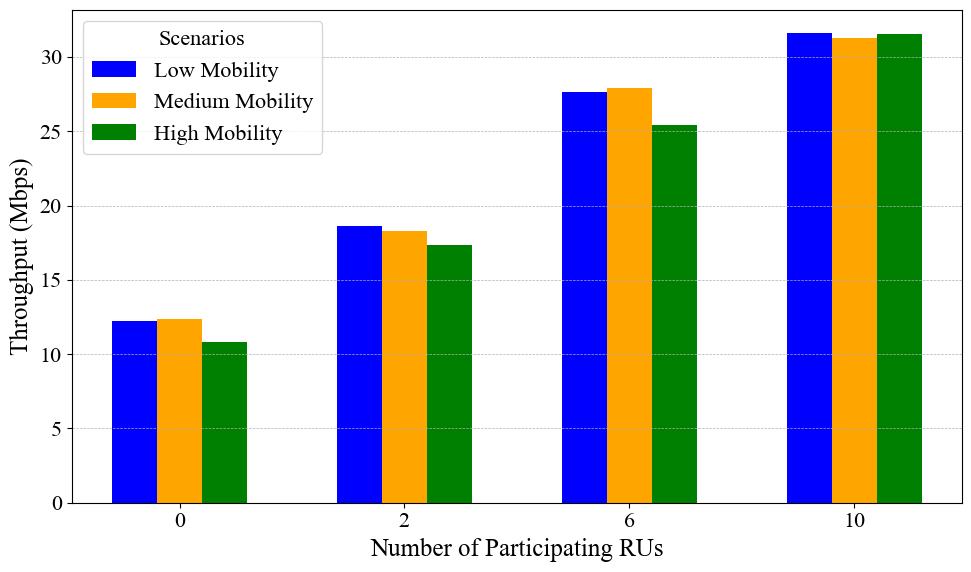}
\caption{Non-coherent MD-MIMO for Uplink with Two UEs. With each choice of number of RUs participating, the throughput is derived by maximizing over various modulation and coding schemes (MCS). It is apparent that with higher number of RUs, the system can handle higher order MCS.} \label{fig:uplink_results}
\end{figure}

As illustrated in Fig. \ref{mult_tx_arry_angl_rx_arry}, 
two UEs attempt to communicate with the receive gNB with the assistance of multiple RUs.
Once each RU has separated the two streams in phase 1, every RU holds a copy of the detected data for each UE.
In this scheme the RUs will decode and forward their copies of their detected data to the receive gNB over the high-capacity phase-2 link.
It then falls upon the receive gNB to consolidate the copies of each UE's detected data arriving from the different RUs.
In this consolidation process, the receive gNB may employ a variety of post-detection combination techniques, such as maximum likelihood, to optimally merge the information.
Figure \ref{fig:uplink_results} presents the throughput achieved by this scheme as the number of participating RUs varies.
In this figure, the throughput is maximized over various modulation and coding schemes (MCS) for each case corresponding to different number of RUs.
In particular, as the size of the virtual array increases, it becomes possible to utilize higher-order MCS, which in turn increases throughput. 
Moreover, it is observed that this architecture exhibits lower sensitivity to mobility effects compared to the transmit virtual array architecture used for downlink.



\section{Conclusion and Outlook}\label{sec:conclusions}

Novel architectures and algorithms for MD-MIMO will be key to further enhancing the application scenarios in NextG wireless systems.

In this article, we discuss various key system design aspects of MD-MIMO including use cases, network architectures, grand challenges, and corresponding enabling technologies. 
Depending on the system design constraints, coherent MD-MIMO and non-coherent MD-MIMO have been introduced. 
Initial performance results from our early studies and prototyping focusing on non-coherent MD-MIMO technologies have been presented to demonstrate performance improvements in both reliability and spectral-efficiencies. 

Overall, we believe MD-MIMO is a promising technology for NextG networks to significantly improve network performance with enhanced deployment flexibility.
 


%

\bibliographystyle{IEEEtran}
\bibliography{IEEEabrv,references.bib}

\end{document}